\documentclass[sigconf,screen]{acmart}

\DeclareMathOperator*{\bigast}{\scalebox{1.8}{\raisebox{-0.2ex}{$\ast$}}}
\title{Uniqueness is Separation}
\author{Liam O'Connor}
\affiliation{
\institution{University of Edinburgh}
  \city{Edinburgh}
  \country{Scotland}
}
\email{l.oconnor@ed.ac.uk}
\author{Pilar Selene Linares Ar\'evalo}
\affiliation{
\institution{University of Melbourne}
\city{Melbourne}
\country{Australia}
}
\email{plinaresarev@student.unimelb.edu.au}
\acmConference{Value Independence in Modern Programming Languages}{March 2023}{Tokyo, Japan}
\author{Christine Rizkallah}
\affiliation{
\institution{University of Melbourne}
\city{Melbourne}
\country{Australia}
}
\email{christine.rizkallah@unimelb.edu.au}
\acmConference{VIMPL}{March 2023}{Tokyo, Japan}
\acmDOI{}
\acmISBN{}
\setcopyright{none}
\begin{abstract}
Value independence is enormously beneficial for reasoning about software systems at scale. These benefits carry over into the world of formal verification. Reasoning about programs algebraically is a simple affair in a proof assistant, whereas programs with unconstrained mutation necessitate much more complex techniques, such as Separation Logic, where invariants about memory safety, aliasing, and state changes must be established by manual proof.

Uniqueness type systems allow programs to be compiled to code that uses mutation for efficiency, while retaining a semantics that enjoys value independence for reasoning. The restrictions of these type systems, however, are often too onerous for realistic software. Thus, most uniqueness type systems include some ``escape hatch'' where the benefits of value independence for reasoning are lost, but the restrictions of uniqueness types are lifted.
To formally verify a system with such mixed guarantees, the value independence guarantees from uniqueness types must be expressed in terms of imperative, mutable semantics. In other words, we ought to express value independence as an assertion in Separation Logic.
\end{abstract}
\begin{document}
\begingroup

\mathchardef\UrlBreakPenalty=10000
\maketitle
\section{Introduction}
 
Uniqueness types \cite{Smetsers1994} allow reasoning about a program as if all data structures in the program are immutable, with all of the benefits that implies, while the actual implementation performs efficient destructive updates to mutable data structures. This is achieved by statically ruling out every program where the difference between the immutable and the mutable interpretations can be observed, by requiring that every mutable value has only one live reference to it at a time. This is called the \emph{uniqueness condition}, which guarantees the non-aliasing necessary to ensure that immutable and mutable semantics coincide. The Cogent language~\cite{jfp} utilises these reasoning benefits to generate purely functional specifications that are easy to reason about and connect them by formal proof to efficient C code that makes use of destructive updates.

The uniqueness condition is a very simple restriction, but it can impose a considerable burden when trying actually to write programs. For example, a simple uniqueness type system would prohibit passing both an array and a reference to one of its elements to a function because of the aliasing this introduces, even if we are only reading from these references and no mutation is involved. Sophisticated data structures, with complex layouts that involve many shared and aliasing references, simply cannot be expressed.

For this reason, most uniqueness type systems include some kind of ``escape hatch'' where the restrictions imposed by the type system can be temporarily suspended. For example, in Rust~\cite{rust}, code in \texttt{unsafe} blocks may violate the uniqueness condition. Cogent similarly includes a foreign function interface (FFI) that allows parts of the program to be written in the unsafe imperative C language. These C components are able to manipulate opaque types that are abstract in the purely functional Cogent components of the system. The Cogent certifying compiler's refinement theorem would then include assumptions that the C components are safe and that they do not violate the uniqueness condition for the Cogent parts of the system. Verification of a mixed Cogent-C system then requires these assumptions to be discharged by manual proof for the C code, as described by \citet{Cheung_OR_22}.
  
This position paper describes the obligations that Cogent imposes about references and mutable stores\footnote{There are other obligations too. See \citet{Cheung_OR_22} for a full list.}, and justifies our belief that Separation Logic~\cite{Reynolds2002} provides the right language to express these obligations and the right proof calculus to discharge them.

\section{Enforcing Uniqueness}
Cogent's static type system is quite sophisticated, and we will not recapitulate it here\footnote{See \citet{jfp} for a full description.}. For our purposes, we are only interested in the dynamic properties that it ensures. 
Cogent's type preservation theorem shows that a function of type $\tau \rightarrow \rho$, will, given a value $v$ of type $\tau$, indeed return a value of type $\rho$. Moreover, Cogent extends this dynamic typing relation for values $v : \tau\ \langle p \rangle$ to include a set of pointers $p$ that can be accessed from the value $v$, called its \emph{heap footprint}\footnote{The actual footprint is slightly more complicated to account for \emph{borrowing}, additionally including a set of shareable, read-only pointers.}. For example, the typing rule for tuple values is:
$$
 \dfrac{ x : \tau_1\ \langle p \rangle \quad                y : \tau_2\ \langle q \rangle \quad                p \cap q = \emptyset               }{ (x, y) : \tau_1 \times \tau_2\ \langle p \cup q \rangle } 
 $$
Observe how these pointer sets are used to enforce that there is no internal aliasing in the structure (the premise $p \cap q = \emptyset$). This is because tuples are not abstract to Cogent and thus any internal aliasing could lead to a violation of the uniqueness condition. For external C-implemented data structures, which are abstract to Cogent, internal aliasing can be permitted.  

This typing relation also provides the information necessary to precisely state the conditions under which a C program will not interfere with the memory safety or value independence guarantees that Cogent enjoys:
Let $\sigma$ denote a \emph{store}, i.e.\ a partial mapping from a pointer $\ell$ to a value $v$. Let $f : \tau \rightarrow \rho$ be a function implemented in C and imported into Cogent. If $f$ is evaluated with an input value $v : \tau\ \langle p \rangle$ and an input store $\sigma$, the return value $f\ v : \rho\ \langle p' \rangle$ and output store $\sigma'$ must satisfy the following three properties for all pointers $\ell$ :

\begin{description}
\item[Leak freedom] $\ell \in p \land \ell \notin p' \Rightarrow \ell \notin \text{dom}(\sigma')$, that is any input reference that was not returned was freed.
\item[Fresh allocation] $\ell \notin p \land \ell \in p' \Rightarrow \ell \notin \text{dom}(\sigma)$, that is every new output reference, not in the input, was allocated in previously-free space.
\item[Inertia] $\ell \notin p \land \ell \notin p' \Rightarrow \sigma(\ell) = \sigma'(\ell)$, that is, every reference not in either the input or the output of the function has not been modified in any way.
\end{description}
Assuming these three properties, it is possible to show that the two semantic interpretations of programs with uniqueness types are equivalent, even if they depend on unsafe, imperative C code. These three conditions are called the \emph{frame conditions}, named after the frame problem from the field of knowledge representation. 

\section{Separation Logic}

The aforementioned frame problem is a common issue that frequently arises in formalisations of stateful processes. Specifically, it refers to the difficulty of \emph{local reasoning}. For example, typical imperative programs lend themselves to axiomatic semantics for verification, the most obvious example being Hoare Logic~\cite{Hoare1969}, which provides a proof calculus for a judgement written $\mu \models \{ \phi \} P \{ \psi \}$. This states that assuming the initial state $\mu$ (which maps variables to values) satisfies an assertion $\phi$, then the resultant state of running $P$ on $\mu$, satisfies $\psi$.  Verification frameworks based on Hoare logic work well for simple programs, but programs that manipulate memory in a heap are tedious and difficult to verify.  Several invariants must be carried around to say that references do not alias, references point to free space, or references point to valid values. This is because the heap is treated as one monolithic structure. Therefore, whenever any part of the heap is updated, every invariant about the heap must be re-established, even if it is independent of the change --- this is the frame problem. 

The Cogent frame conditions state that any function, including those implemented in C, does not affect any part of the heap except those it is permitted (by virtue of the references it received) to modify, thus ensuring that our invariants are preserved for all other parts of the program. While such a presentation of the frame conditions is fine for the automatic proofs generated for Cogent code, presenting such proof obligations directly in terms of heaps and pointers remains tedious and difficult when verifying the C components of the system~\cite{Cheung_OR_22}. This is particularly the case when our invariants must be initially broken and only re-established later.

To make this cleaner, we turn instead to the Separation Logic of \citet{Reynolds2002}. Separation Logic is a variant of Hoare Logic that is specifically designed to accommodate programming with references and aliasing. In addition to the state $\mu$ of Hoare Logic, we have a mutable store $\sigma$, and the following additional assertions:
\begin{itemize}
\item A special assertion $\mathbf{emp}$, which states that the store is empty, i.e $\mu, \sigma \models \mathbf{emp}$ if and only if $\text{dom}(\sigma) = \emptyset$.
\item A binary operator $\mapsto\ : \ell \times v$ , which states that the store is defined at exactly one location, i.e. $\mu, \sigma \models \ell \mapsto v$ if and only if $\text{dom}(\sigma) = \{ \ell \} \land \sigma(\ell) = v$.
\item A \emph{separating conjunction} connective $\phi \ast \psi$, which says that the store $\sigma$ can be split into two disjoint parts $\sigma_1$ and $\sigma_2$ where $\mu, \sigma_1 \models \phi$ and $\mu, \sigma_2 \models \psi$.
\item A \emph{separating implication} connective $\phi -\!\!\!\ast\ \psi$, which says that extending the store with a disjoint part that satisfies $\phi$ results in a store that satisfies $\psi$.
\end{itemize}
Crucially, Separation Logic includes the \emph{frame rule}, its own solution to the frame problem, where an unrelated assertion $\phi_r$ can be added to both the pre- and the post-condition of a given program in a separating conjunction:
\begin{center}$
 \dfrac{ \{\phi\}\ P\ \{\psi\} }{ \{\phi \ast \phi_r\}\ P\ \{\psi \ast \phi_r\} } 
 $\end{center}
This allows much the same local reasoning that we desired before: The program $P$ can be verified to work for a store that satisfies $\phi$, but otherwise contains \emph{no other values}. Then that program may be freely used with a larger state, and we automatically learn, from the frame rule, that any unrelated bit of state cannot affect and is not affected by the program $P$.

Separation Logic makes expressing our frame conditions much simpler. Given a program $P$ with an input set of pointers $p$ and output set of pointers $p'$, we express all three conditions as a single triple:
\begin{center}$
   \left \{ \bigast_{\scriptstyle \ell \in p} \exists v.\ \ell \mapsto v \right \} P \left \{ \bigast_{\ell \in p'} \exists v.\ \ell \mapsto v \right \} 
$\end{center}
We sketch of a proof that this implies the frame conditions listed above. Assume an input store $\sigma$. Split $\sigma$ into disjoint stores $\sigma_1$ and $\sigma_2$ such that $\sigma_1 \models  \bigast_{\ell \in p} \exists v.\ \ell \mapsto v\ \ (*)$ . Let the output store of running $P$ with $\sigma_1$ be $\sigma_1'$. Note that by the triple above, we have that $\sigma_1' \models  \bigast_{\ell \in p'} \exists v.\ \ell \mapsto v\ \ (*\!*)$.
Using the frame rule, we know that the output of running $P$ with the full store $\sigma$ is $\sigma' = \sigma_1' \cup \sigma_2$ where $\text{dom}(\sigma_1') \cap \text{dom}(\sigma_2) = \emptyset$.
\begin{description}
\item[Leak freedom] For any arbitrary location $\ell$, if $\ell \in p$ but $\ell \notin p'$ then we must show that $\ell \notin \text{dom}(\sigma')$. As $\ell \in p$, we know from $(*)$ that $\ell \in \text{dom}(\sigma_1)$ and, as they are disjoint, $\ell \notin \text{dom}(\sigma_2)$ . Therefore, the only way for $\ell \in \text{dom}(\sigma')$ to be true is if $\ell \in \text{dom}(\sigma_1')$, but as $\text{dom}(\sigma_1') = p'$ from $(*\!*)$, we can conclude that $\ell \notin \text{dom}(\sigma')$.
\item[Fresh allocation] If $\ell \notin p$ but $\ell \in p'$ then we must show that $\ell \notin \text{dom}(\sigma)$. We have from $(*\!*)$ that $p' = \text{dom}(\sigma_1')$, and hence $\ell \in \text{dom}(\sigma_1')$. As they are disjoint, $\ell \notin \text{dom}(\sigma_2)$ so the only way for $\ell \in \text{dom}(\sigma)$ to be true is if $\ell \in \text{dom}(\sigma_1)$. But, as we know that $\text{dom}(\sigma_1) = p$ from $(*)$ and $\ell \notin p$, we can conclude that $\ell \notin \text{dom}(\sigma)$.
\item[Inertia] If $\ell \notin p$ and $\ell \notin p'$, then we can conclude from $(*)$ that $\ell \notin \text{dom}(\sigma_1)$ and from $(*\!*)$ that $\ell \notin \text{dom}(\sigma_1')$. If $\ell \in \text{dom}(\sigma_2)$, then $\sigma(l) = \sigma_2(l) = \sigma'(l)$, thanks to the frame rule as shown above. If $\ell \notin \text{dom}(\sigma_2)$, then $\ell \notin \text{dom}(\sigma)$ and $\ell \notin \text{dom}(\sigma')$.
\end{description}
This presentation also enables the reuse of existing successful frameworks for C verification~\citep{Callum2018} based on Separation Logic. 
\section{Conclusion}
We believe that this elegant formulation of frame conditions is evidence of a deep connection between uniqueness types and Separation Logic. Type systems and program logics are both tools for formal reasoning, and we run the risk of reinvention if we do not realise the connections between them. These connections could be the basis of a futuristic language combining refinement types~\cite{refinementtypes} and uniqueness types, compiling down to efficient imperative code where type-based assertions are translated into assertions in Separation Logic, enabling seamless integration and verification with hand-coded low-level extensions. Such a language would lower the barriers of tedium that plague existing verification efforts, and broaden the horizons for verified software development.
\bibliography{refs}

\begin{thebibliography}{8}
\providecommand{\natexlab}[1]{#1}
\providecommand{\url}[1]{\texttt{#1}}
\expandafter\ifx\csname urlstyle\endcsname\relax
  \providecommand{\doi}[1]{doi: #1}\else
  \providecommand{\doi}{doi: \begingroup \urlstyle{rm}\Url}\fi

\bibitem[Bannister et~al.(2018)Bannister, H{\"o}fner, and Klein]{Callum2018}
Callum Bannister, Peter H{\"o}fner, and Gerwin Klein.
\newblock Backwards and forwards with separation logic.
\newblock In Jeremy Avigad and Assia Mahboubi, editors, \emph{Interactive
  Theorem Proving}, pages 68--87, Cham, 2018. Springer International
  Publishing.
\newblock ISBN 978-3-319-94821-8.

\bibitem[Bierman et~al.(2010)Bierman, Gordon, Hri\c{t}cu, and
  Langworthy]{refinementtypes}
Gavin~M. Bierman, Andrew~D. Gordon, C\u{a}t\u{a}lin Hri\c{t}cu, and David
  Langworthy.
\newblock Semantic subtyping with an smt solver.
\newblock \emph{SIGPLAN Not.}, 45\penalty0 (9):\penalty0 105–116, sep 2010.
\newblock ISSN 0362-1340.
\newblock \doi{10.1145/1932681.1863560}.
\newblock URL \url{https://doi.org/10.1145/1932681.1863560}.

\bibitem[Cheung et~al.(2022)Cheung, O'Connor, and Rizkallah]{Cheung_OR_22}
Louis Cheung, Liam O'Connor, and Christine Rizkallah.
\newblock Overcoming restraint: Composing verification of foreign functions
  with cogent.
\newblock In \emph{Proceedings of the 11th ACM SIGPLAN International Conference
  on Certified Programs and Proofs}, CPP 2022, page 13–26, New York, NY, USA,
  2022. Association for Computing Machinery.
\newblock ISBN 9781450391825.
\newblock \doi{10.1145/3497775.3503686}.

\bibitem[Hoare(1969)]{Hoare1969}
C.~A.~R. Hoare.
\newblock An axiomatic basis for computer programming.
\newblock \emph{Commun. ACM}, 12\penalty0 (10):\penalty0 576–580, 1969.
\newblock ISSN 0001-0782.
\newblock \doi{10.1145/363235.363259}.
\newblock URL \url{https://doi.org/10.1145/363235.363259}.

\bibitem[Klabnik and Nichols(2018)]{rust}
Steve Klabnik and Carol Nichols.
\newblock \emph{The Rust Programming Language}.
\newblock No Starch Press, USA, 2018.
\newblock ISBN 1593278284.

\bibitem[O'Connor et~al.(2021)O'Connor, Chen, Rizkallah, Jackson, Amani, Klein,
  Murray, Sewell, and Keller]{jfp}
Liam O'Connor, Zilin Chen, Christine Rizkallah, Vincent Jackson, Sidney Amani,
  Gerwin Klein, Toby Murray, Thomas Sewell, and Gabriele Keller.
\newblock {Cogent}: uniqueness types and certifying compilation.
\newblock \emph{Journal of Functional Programming}, 31:\penalty0 {e25}, 2021.
\newblock \doi{10.1017/S095679682100023X}.

\bibitem[Reynolds(2002)]{Reynolds2002}
J.C. Reynolds.
\newblock Separation logic: a logic for shared mutable data structures.
\newblock In \emph{Proceedings 17th Annual IEEE Symposium on Logic in Computer
  Science}, pages 55--74. IEEE, 2002.
\newblock \doi{10.1109/LICS.2002.1029817}.

\bibitem[Smetsers et~al.(1994)Smetsers, Barendsen, van Eekelen, and
  Plasmeijer]{Smetsers1994}
Sjaak Smetsers, Erik Barendsen, Marko van Eekelen, and Rinus Plasmeijer.
\newblock Guaranteeing safe destructive updates through a type system with
  uniqueness information for graphs.
\newblock In Hans~J{\"u}rgen Schneider and Hartmut Ehrig, editors, \emph{Graph
  Transformations in Computer Science}, pages 358--379, Berlin, Heidelberg,
  1994. Springer Berlin Heidelberg.
\newblock ISBN 978-3-540-48333-5.
\newblock \doi{10.1007/3-540-57787-4_23}.

\end{thebibliography}
\bibliographystyle{plainnat}
\end{document}